\documentclass[epsfig]{epl}

\newlength{\GraphicsWidth}
\newcommand\comment[1]{\textsc{{ 1}}}

\newcommand{\bv}{\mathbf{v}}

\newcommand{\be}{\begin{equation}}
\newcommand{\ee}{\end{equation}}

\begin{document}
%\euro{}{}{}{}
%\Date{}

%\shorttitle {Boltzmann equation for soft and hard spheres}

\title{
%Boltzmann equation for inelastic soft and hard spheres}
The rich behavior of the Boltzmann equation for
dissipative gases}

\author{M.H. Ernst\inst{1}, E. Trizac\inst{2} and A. Barrat\inst{3}}
\institute{
\inst{1}{Theoretische Fysica, Universiteit
Utrecht, Postbus 80.195, 3508 TD Utrecht (The Netherlands)} \\
\inst{2}{Theoretical Biological Physics,
UC San Diego, La Jolla CA  92093 (USA) and LPTMS (UMR CNRS 8626),
Universit\'e Paris-Sud, 91405 Orsay (France)} \\
\inst{3}{LPT (UMR CNRS 8627),
Universit\'e Paris-Sud, 91405 Orsay (France)}
}

%\pacs{
%}

\maketitle

\begin{abstract}
Within the framework of the homogeneous non-linear Boltzmann
equation, we present a new analytic method, without the intrinsic
limitations of existing methods, for obtaining asymptotic
solutions. This method permits extension of existing results for
Maxwell molecules and hard spheres to large classes of particle
interactions, from very hard spheres to softer than Maxwell
molecules, as well as to more general forcing mechanisms, beyond
free cooling and white noise driving. By combining this method with
numerical solutions, obtained from the Direct Simulation Monte
Carlo (DSMC) method, we study a broad class of models relevant for
the dynamics of dissipative fluids, including granular gases. We
establish a criterion connecting the stability of the
non-equilibrium steady state to an exponentially bound form for the
velocity distribution $F$, which varies depending on the forcing
mechanism. Power laws arise in marginal stability cases, of which
several new cases are reported. Our results provide a minimal
framework for interpreting large classes of experiments on driven
granular gases.
\end{abstract}

%\keywords{}

Granular materials are ubiquitous in natural phenomena and widely
used in industrial processes. Yet, a fundamental understanding of
their properties still challenges scientists
\cite{Jaeger,PoschelBrill}. Although a large number of particles
is involved, a thermodynamic-like description remains elusive and
has been thwarted by energy losses through internal dissipation,
and energy supplies from non-thermal sources. Granular media are
therefore intrinsically far from equilibrium, and a paradigm for
open dissipative systems. One can distinguish two types of
granular fluids \cite{Jaeger}: {\it quasi-static} flows
(solid-like), where gravity, friction and stress distributions
control the dynamics, and {\it rapid granular} flows (gas-like),
where the dynamics is mostly governed by ballistic motion in
between inelastic collisions. This letter deals with
granular gases.

Even dilute granular gases differ from their molecular counterparts
to a significant extent~\cite{Jaeger,PoschelBrill}, and their
particle velocity distribution is emblematic of the difficulties of
a statistical description.
%From an experimental point of view, the
%situation may appear confusing at first sight. Although
The experimentally measured velocity distribution $F({\bf v})$
generically deviates from the Maxwell-Boltzmann behavior, and its
functional form --often fitted by stretched exponentials-- depends
on material property, on the specific geometry considered and on
the forcing mechanism which compensates the collisional loss of
energy \cite{Exp,Aran-Olaf,Prevost,Aranson,Melby}. A similar
picture emerges from numerical simulations and analytical studies
\cite{Aranson,Gold,Brey,Andrea,stretch-tails,MS,Cafiero,Swinney,Santos,Maxwell,ME-et-al,Math,BBRTvW,
Pias,ben-Avb,Zip,vanZon,Haifa,EBN-Machta}, where in addition to the
more common stretched exponential behavior, power-law distributions
have also been reported \cite{Maxwell,ME-et-al,EBN-Machta}. This
rather vast but disparate body of knowledge calls for
rationalization.

 Our goal is to propose a general and unified framework for
interpreting  experimental results, as well as to generalize
existing theoretical results in two new and important directions:
(i) to more general repulsive interactions, parametrized by an
exponent $\nu$ , which include the known results for Maxwell
molecules $(\nu=0)$ and hard spheres ($\nu =1$), and (ii) to more
general forcing mechanisms, such as nonlinear negative friction
forces, parametrized by an exponent $\theta$, including free cooling
($\theta =1$), as well as the white noise driving force. In
addition, since the Maxwell-Boltzmann distribution seems to have no
analog in steady states of dissipative gases, we want to identify
the generic trends for the velocity distribution and study its
properties in this framework. To this aim, we develop a global and
quantitative cartoon of the different effects at work, that lead,
upon changing certain key parameters, to the wide range of
behaviors alluded to above.

In this letter, we develop a new analytic method for determining
deviations from Gaussian behavior, that is free of the intrinsic
limitations of previous approaches. For instance, the known Fourier
transform method for obtaining high energy power law tails
\cite{Bobylev,Maxwell,ME-et-al} is intrinsically restricted to
Maxwell molecules, where the nonlinear collision term is a
convolution product. Moreover, the method for obtaining stretched
exponential tails in \cite{stretch-tails,ME-et-al} was based on a
one-stroke asymptotic approximation, whereas our new method
provides a systematic approximation scheme yielding large
sub-leading corrections. We also show that the large velocity tail
of $F(\bv)$ is characterized by an exponent $b$, which governs the
stability of the non-equilibrium steady state. When this state is a
stable fixed point of the dynamics, $b>0$ and $F$ is of stretched
exponential form $\sim \exp[-v^b]$. We also emphasize the relevance
of subdominant corrections that must be taken into account for a
comparison of numerical or experimental data with the theory. For
cases of marginal stability where $b$ vanishes, new classes of
power law tails $v^{-s}$ appear.
An immediate experimental consequence is
that such power law tails will be elusive, since they can only
arise when material properties are fine tuned. Small variations
will either drive the system into an unstable state, or into a
stable state with a stretched exponential tail.
%The stretching exponent $(b)$ and power law exponent depend
%sensitively on $\nu$ and $\theta$.

We use the simplest model for rapid granular flows, i.e. the
nonlinear Boltzmann equation, that corresponds to a regularly
'randomized' gas (by shaking, fluidization, vibrational techniques,
or bulk driving through alternating electric fields). The dynamics
is described as a succession of uncorrelated inelastic binary
collisions, modelled by soft spheres with collision frequency, $g
\varsigma (g,\vartheta) \sim g^\nu |\widehat{\mathbf g}\cdot
\bf{n}|^\sigma$, and a coefficient of normal restitution $\alpha<1$.
The collision
law $(\bv_1,\bv_2)\to (\bv'_1,\bf v'_2)$ reads: $\bv'_{1} = \bv_{1}
- p(\bf{g}\!\cdot \!
\bf{n}) \bf{n}$ where $ \bf{g} \equiv \bv_1-\bv_2$, $ p= 1-q
= \frac{1}{2}(1+\alpha)$ and $\bf{n}$ is a unit vector parallel to
the impact direction connecting particles 1 and 2. The expression
for $\bv'_2$ follows from momentum conservation. Momentum
transfer at a distance can also take place through long range interactions
\cite{Aran-Olaf,Aranson}.
Collisions dissipate kinetic energy at a rate $\propto {1\over
2}(1-\alpha^2)= 2pq$. The quantity $\varsigma (g,\vartheta) $ is
the differential scattering cross section, where exponent $\sigma$
models the angular dependence, and $\mathbf{\widehat{g}} = {\bf g}
/g $ is a unit vector, parallel to $\bf{g}$. For elastic particles
interacting via a soft sphere potential $U(r)\propto r^{-a}$, one
has $\nu = 1-2(d-1)/a$, where $d$ is the space dimension. The
exponents $\nu= \sigma=1$ model ordinary hard-sphere behavior
$(a\to \infty)$, and $\nu=0$ defines so-called Maxwell models. The
hallmark of Maxwell interactions is a collision frequency
independent of the relative velocity of the colliding pair. In
Ref.\cite{Aranson} this property has been found experimentally, and
explained theoretically for experiments on non-magnetic particles
in a viscous fluid, and magnetic grains with dipolar interactions
in air in a 2-D geometry where $U(r)=r^{-2}$. Because of this
hallmark property, Maxwell models are particularly convenient for
analytical purposes. Note also that the importance of material
properties has barely been taken into account in previous studies.
Here we do so in a necessarily schematic but physically relevant
manner through the parameters $\nu$, $\sigma$ and $\alpha$, that 
encode material properties.

The study of the homogeneous systems is not
only a useful starting point, it is also relevant to experiments
with bulk driving \cite{Prevost,Aranson}, and a significant
fraction of the experiments %listed
\cite{Exp,Aran-Olaf,Aranson,Prevost,Melby} has been carried out on
spatially uniform systems. The homogeneous non-linear Boltzmann reads
%\begin{eqnarray}
\begin{equation}
\partial_t  %\!\!\!
F({\bv},t) + {\cal F} F=
\int' d{\bv}_1 d{\bv}_2 d\bf{n}\,  g^\nu |\widehat{\bf{g}}\cdot{\bf{n}}|^\sigma
%\nonumber\\ &&
F({\bv}_1,t) F({\bv}_2,t) [\delta(\bv-\bv'_1)-
\delta(\bv-\bv_1)] ,
\label{eq:Boltz}
\end{equation}
where the r.h.s defines the collision kernel $I(\bv|F)$ with
the usual gain/loss structure and
Dirac functions to enforce the collision law;
\smash{$\int'$} indicates that the $\bf{n}$
-integration is an average over the pre-collision hemisphere,
$\mathbf{g}\cdot\bf{n}\leq0$, with a uniform weight normalized to
unity. The forcing term ${\cal F}F$ represents the energy
supply.
%, working against inelastic dissipation.
We consider here the generic form
\be
{\cal F}F \,=\,  \gamma
\,\partial_{\bv}
\cdot({\bv} v^{\theta-1} F) \,-\, D \, \partial_{{\bv}}^2 F
\label{eq:forcing}
\ee with either ($\gamma=0$, $D>0$) or ($\gamma >0$, $D=0$),
corresponding respectively to {\it stochastic} White Noise (WN), or
{\it deterministic} nonlinear Negative Friction (NF), where $\theta
\geq 0$ is a continuous exponent. The change in velocity due to
the latter force is proportional to $ v^{\theta -1} {\mathbf{v}}$.
For instance, for $\theta >1 $ it injects selectively energy in the
high energy tail of $F(v)$. A physical realization of NF with
$\theta>1$ is provided by an electrically forced gas with a charge
distribution. Energy is predominantly injected in the particles
with a high charge, which in turn reach larger typical velocities.
Recently a somewhat different device with a similar property has
been proposed in Ref.\cite{EBN-Machta}. The special case $\theta=1$
corresponds to the Gaussian thermostat \cite{MS}, allowing to study
the long time scaling regime of an unforced system (so-called free
cooling), while $\theta=0$ models gliding friction, also called
gravity thermostat \cite{MS}. It corresponds to a force of constant
strength acting in the direction of the velocity. WN forcing has
been frequently used in analytical and numerical studies
\cite{stretch-tails,MS,Swinney,BBRTvW,vanZon}, and more importantly,
was shown to be experimentally relevant for two dimensional
experiments with energy injected from a vibrating rough
plate~\cite{Prevost,Melby}. In such cases individual grains undergo
random changes in direction, that are mimicked by white noise
forcing. Existing experimental evidence also shows that stochastic
forcing is frequently a valuable first approximation
\cite{Aran-Olaf,Swinney}. We therefore have a restricted 'phase
space' to explore:($\nu,\sigma,\alpha$) for WN and
($\nu,\sigma,\alpha,\theta$) for NF. Here $\gamma$ and $D$ are
irrelevant constants that set the energy scale.

While the dynamics defined by (\ref{eq:Boltz}) and
(\ref{eq:forcing}) always admits a non-equilibrium steady state,
the first question to address is the stability of this steady
state. Our analysis assumes a rapid approach of $F$ to a scaling
form $F(v,t) = (1/v_0(t))^d f(v/v_0(t))$ (for proofs, see
\cite{Math}),  where the r.m.s.
velocity $v_0$ is defined through the granular temperature, $\int
d{\bf v} v^2 F(v,t) = {1\over 2} d v_0^2(t)$, and
Eq.(\ref{eq:Boltz}) implies:
\begin{eqnarray} \label{eq:v0}
&d v_0^2/dt \propto 1-\left(v_0(t)/v_0(\infty)\right)^{2b}
\quad (b_{\rm{WN}}=1+\nu/2)&  \nonumber\\
& d v_0/dt \propto  v_0^\theta (t) \left [
1-\left(v_0(t)/v_0(\infty)\right)^{b} \right ]
\quad (b_{\rm{NF}}=1+\nu-\theta)&
\end{eqnarray}
The solution $v_0$ therefore approaches the fixed point
$v_0(\infty)$ if $b>0$, and conversely moves away from
$v_0(\infty)$ if $b<0$, in which case the steady state is an
unstable fixed point of the dynamics.

In stable steady states  the integral equations for $f(c)$ can be
derived from (\ref{eq:Boltz}), and read for WN and NF driving
respectively:
\begin{eqnarray} \label{eq:int}
&I(c|f) = -\frac{D}{(v_0(\infty))^{2b}} \partial^2_{{\bf c}} f =
-\frac{\lambda_2}{4d}
\langle\langle g^{\nu+2}\rangle\rangle \partial^2_{{\bf c}} f &
\nonumber \\
&I(c|f) = \frac{\gamma}{(v_0(\infty))^b} \partial_{{\bf c}}\!\cdot
\!(\widehat{{\bf c}} c^\theta f) =  \frac{\lambda_2
\langle\langle g^{\nu+2}\rangle\rangle}{4 \langle c^{\theta
+1}\rangle}{\partial}_{{\bf c}}\!\cdot \! (\widehat{{\bf c}}
c^\theta f)&
\end{eqnarray}
The phenomenological constants $D$ and $\gamma$ have been
eliminated with the help of the steady state conditions in
(\ref{eq:v0}). The averages $\langle\langle \cdots\rangle\rangle$
and $\langle \cdots\rangle$ are respectively taken over the unknown
$f(c_1)f(c_2)$ and $f(c)$, and $g =|\bf{ c}_1 -\bf{c}_2|$. Moreover
$\lambda_2 =2pq \beta_{\sigma+2}$ and $\beta_s \equiv \int
d{\bf{n}} \:{\bf |\widehat{a} \cdot \bf{n}}|^s /\int {d}\bf{n}$ is
an average of $|\cos\vartheta|^s $ over a $d$-dimensional solid
angle.

Next we discuss the new criterion relating stretching exponents to
stability. From the one stroke asymptotic approximation of
\cite{stretch-tails} an interesting feature already emerges:
in the stability region ($b>0$), $f(c)$ is of stretched exponential
form with an {\it a priori} known exponent $b$  from (\ref{eq:v0}),
while in cases of marginal stability ($b
\to 0^+$), $f \sim c^{-a}$ is of power law type with an {\it a
priori} unknown exponent $a$. As expected, $b$ decreases with
$\nu$, since a tail particle with velocity $c \gg 1$ suffers
collisions at a rate $c^\nu$; the slower the rate, the slower the
particle redistributes
its energy over the thermal range $c % \lesssim
\stackrel{<}{\sim} 1 $, which results in
an increasingly overpopulated high energy tail. When $\nu$ is
further decreased so that $b$ changes sign, the tail is no longer
able to equilibrate with the thermal ``bulk'', and the system
cannot sustain a steady state. A similar intuitive picture may be
developed with respect to $\theta$ in the NF cases. While
$\theta=1$ is equivalent to a linear rescaling of
velocities, a nonlinear rescaling with $\theta>1$ selectively puts more
energy in the tail particles, thus leading to an increased
overpopulation, as $b$ decreases (see (\ref{eq:v0})). The reverse
scenario applies for $\theta < 1$.

\begin{figure}[t]
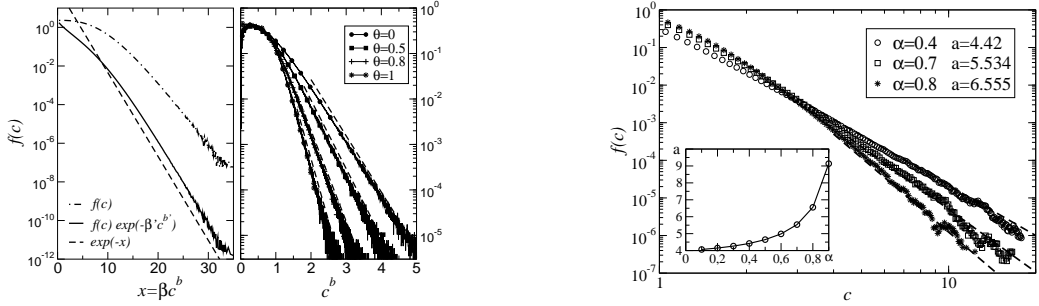

\includegraphics[height=4cm,angle=0]{fig1-epl.eps}
\hspace{2cm}
\includegraphics[height=4cm,angle=0]{fig2-epl.eps}
\caption{Comparison of the velocity distribution obtained from Monte Carlo
simulations with the asymptotic predictions. Left: WN forcing for
$d=2$, $\alpha=1/2$, $\sigma=\nu= -1/2$, where $b=3/4,b'=9/16,\chi
=0$. DSMC data are plotted as $f(c)$ (dashed-dotted line) and
$\exp[-\beta' c^{b'}] f(c)$ (solid line) vs $x= \beta c^b$, and
compared with the theoretical prediction $e^{-x}$ (dashed line).
Right: NF forcing with $d=1$, $\alpha =0,\nu =1/2$. The DSMC data
for $f(c)$ vs $c^b$ are compared at various $\theta$ with the
 prediction $\exp[-\beta c^b]$ (dashed lines).
$\beta,\beta'$ have been measured in the DSMC simulations from
their definition (\ref{eq:beta}).
 Fig.2: WN-driven system at
stability threshold ($d=2,\nu=-2,\sigma=1$) for different values of
the restitution coefficient $\alpha$. The dashed lines correspond
to the theoretically predicted power-law tails $1/c^a$. The inset
shows the algebraic exponent obtained through the numerical
solutions of Eq.(\ref{TE}). }
\label{fig:1}
\end{figure}

To be more specific, the collision operator $I$ in (\ref{eq:Boltz})
for $c\gg 1$, acting on exponentially bound functions $f
\sim \exp[-\beta c^b]$, reduces to a linear operator on $f$,
because $\bf{g} \simeq \bf{c}_1$ and ${ \bf{c}}'_1 \simeq
{\bf{c}}_1 - {p} (\bf{c}_1 \cdot \bf{n}) \bf{n}$, yielding the
dominant behavior of Ref.\cite{stretch-tails}, and a new important
subleading correction~\cite{appear},
\be \label{eq:I-asympt}
I_{\infty}(c|f) =  -\beta_\sigma c^\nu [1 -{\mathcal{K}}_\sigma
c^{-b(\sigma+1)/2}] f .
\ee
Here ${\mathcal{K}}_\sigma $ is non-vanishing, except for 1-D. Its
explicit form  contains averages $\langle
\langle g^s \rangle \rangle $ and $ \langle c^s \rangle $ \cite{appear}.
The leading term is a direct generalization of the standard methods
of Ref.\cite{stretch-tails} for inelastic hard spheres ($\nu =1 $)
and Maxwell molecules ($\nu =0 $). Eq.(\ref{eq:I-asympt}) allows us
to calculate  for $ c \gg 1$ sub-leading multiplicative corrections
to $f(c)$ of the form,
\be \label{eq:f-asympt}
\ln f(c) \sim -\beta c^b +\beta' c^{b'} + \chi \ln c +{\mathcal{O}}(1),
\ee
where $ b>b'>0$. By inserting
(\ref{eq:I-asympt})-(\ref{eq:f-asympt}) in (\ref{eq:int}), and
equating exponents and coefficients of the leading  and sub-leading
terms, we recover the $b-$values in (\ref{eq:v0}), and determine
$\beta, b', \beta', \chi$. We only list the values for $\sigma = 1$
(for further details, see \cite{appear}):
\be \label{eq:beta}
(\beta b)_{\rm WN} =  \sqrt{d(d+1)/(pq
\langle\langle g^{\nu+2} \rangle\rangle) }
\ ; \ (\beta b)_{\rm NF} =  (d+1)\langle
c^{\theta+1}\rangle /(pq \langle\langle g^{\nu+2}\rangle\rangle)
\ee
which includes the special cases considered in Ref.
\cite{stretch-tails}. Regarding the sub-leading  quantities
we find for $\sigma=1 $, all values of $q= \frac{1}{2} (1-\alpha)$
and $d$,
\begin{eqnarray} \label{eq:chi}
\beta'=0, & \chi = &-\textstyle{\frac{1}{4}} \nu -(d-1)q^2/2(1-q^2)
\quad (\mbox{WN}) \nonumber \\
\beta'=0, & \chi = &-\theta + (d-1)q^2/(1-q^2)
\quad (\mbox{NF}) .
\end{eqnarray}
For $\sigma>1 $ we find sub-leading results similar to (\ref{eq:chi})
with $\chi_{\rm WN} = \frac{1}{2} (1-d) - \frac{1}{4} \nu $ and
$\chi_{\rm NF} = 1-d -\theta$. For $\sigma < 1$ we obtain $b'=
\frac{1}{2} b(1-\sigma), \chi=0, \beta' \propto \beta
{\mathcal{K}}_\sigma $, both for WN and NF with corresponding
values for $b$ and $\beta$. A striking example of the importance of
the new sub-leading corrections is shown in Fig.\ref{fig:1}(left)
for a model with $\sigma = - 1/2$. The numerical results are
obtained from the Direct Simulation Monte Carlo technique (DSMC),
which offers an efficient algorithm to solve the nonlinear
Boltzmann equation \cite{Bird}. The dashed-dotted curve in Fig.
\ref{fig:1}(left) shows that the simulated $c-$values are not large
enough to agree with the asymptotic predictions. However, the solid
curve shows a striking agreement with the theory. This demonstrates
the importance of the sub-leading corrections, which allow to
extend the validity of the asymptotic theory to much smaller
$c-$value, which are likely to be helpful in experimental
verification. Such corrections are generic in dissipative gases for
$d \leq 2$, and more pronounced when $\sigma<1$. In 1D they are
absent (see Fig.\ref{fig:1}(right)).

Next we analyze the integral equation (\ref{eq:int}) at the
stability threshold ($b=0$), where power law tails, $f(c) \sim
c^{-a}$, are to be expected. The asymptotic approximations, made in
(\ref{eq:I-asympt}), are no longer valid because of the slow decay
of $f(c_2)$ in $I(c|f)$ at large velocities. Here we proceed as
follows. As far as large velocities are concerned ($c \gg 1$), the
thermal range of $f$ may be viewed to zeroth approximation as a
Dirac delta function $\delta({\bf c})$, carrying all the mass of
the distribution. We therefore write $f{(\bf{c})} =
\delta{(\bf{c})}+h({\bf{c}})$, where $h$ represents the high energy tail, of
negligible weight compared to the thermal bulk. This allows us to
linearize the collision operator:
\be
I({\bf{c}}|\delta+h) =- \Lambda h(c) + {\mathcal{O}}(h^2) ,
\label{eq:Lambda}
\ee
where we have used that $\delta(\bf{c})$ makes the nonlinear
collision term vanish [$I(\bf{c}|\delta)=0$]. Hence,
(\ref{eq:Lambda}) defines the linearized collision operator
$\Lambda$.  Its eigenfunctions are of power law type, and
particularly suited to describe the algebraic decay of $f$ in cases
of marginal stability. For the construction of $\Lambda$, which
differs from its self-adjoint $\Lambda^\dagger$, and of its
eigenfunctions and eigenvalues we refer to \cite{appear}. For the
special case of Maxwell molecules the eigenvalues of the linearized
collision operator have been calculated in
Refs.\cite{Bobylev,Maxwell,ME-et-al} with the help of the Fourier
transform method, which can not be applied to non-Maxwell type
interactions. The essential observation in deriving the spectral
properties is that $\Lambda$, applied to any power $c^a$ generates
a term $\propto c^{a+\nu}$. This property allows us to calculate
the eigenvalues $\lambda_s$, and the corresponding left
eigenfunctions, $\Lambda^\dagger c^s =$ $\lambda_s c^{s+\nu}$, and
right ones, $\Lambda c^{-s-d-\nu} =$ $\lambda_s c^{-s-d}$. These
spectral properties, including
\be \label{eq:eigenvalue}
\lambda_s \!=\! \beta_\sigma \left\{ 1 - _2\!F_1\,\left(
\textstyle{-\frac{s}{2},\frac{\sigma+1}{2};
\frac{\sigma +d}{2}} \,|\,1-q^2\right)\right\} -p^s
\beta_{s+\sigma},
\ee
are new results.  Here $_2\!F_1 $ is a hypergeometric function,
$\lambda_s $ is well-defined for $s
\geq 0$, and $\lambda_0=0$ is an isolated point of the spectrum.
The eigenvalue depends only on the angular exponent $\sigma $, and
not on the energy exponent $\nu $. Moreover, for $s=2$ one has
$\lambda_2 = 2 pq
\beta_{\sigma+2}$ (see Eq.(\ref{eq:int})). Consequently, in the
threshold cases ($b=0$, which fixes the exponent $\nu$ at threshold
to $\nu_{\rm WN}=-2$ or $\nu_{\rm NF}=\theta-1$) the asymptotic
form of $f(c)$ is given by the right eigenfunction that satisfies
the integral equation (\ref{eq:int}). Substitution yields the
transcendental equations:
\vspace{-2mm}
\begin{eqnarray} \label{TE}
\lambda_s  &=& \textstyle{\frac{1}{4d}}s(s+d-2) \lambda_2 \qquad({\rm WN})\nonumber \\
 \lambda_s &=& \textstyle{\frac{1}{2}} s \lambda_2
\langle\langle g^{\theta+1}\rangle/\langle
c^{\theta+1}\rangle \equiv \textstyle{\frac{1}{2}}s\lambda_2
\Gamma(\theta) \: ({\rm NF})
\end{eqnarray}
Their largest root $s^*$ determines the exponent of the
corresponding high energy tail, $ f(c) \sim 1/c^{s^*+d+\nu}$ for
the different driving forces at their stability threshold ($b=0$).

For the Gaussian thermostat ($\theta=1$, $\Gamma(1)=1$) the
threshold model ($b_{\rm NF}= \nu_{\rm NF} =0$) is the well-studied
Maxwell model with a high energy tail, $f(c)\sim 1/c^{s^*+d}$,
first discovered  in Ref.\cite{Maxwell,ME-et-al}. In the 1-D case
(where $\beta_s=1$) the eigenvalue (\ref{eq:eigenvalue}) simplifies
to $
\lambda_s =1-q^s-p^s $, and (\ref{TE}) can be solved analytically
for WN, yielding $s^*(d=1)=3$ and $f(c)
\sim 1/c^{s^*+d-2} \sim 1/c^2$, as well as for NF $(\theta=1)$,
yielding $s^*_{\rm NF}(d=1,\theta=1) =3$ and $f(c)\sim 1/c^{s^* +d
+\theta-1} \sim 1/c^4$. For all remaining ($d,\theta$) values,
(\ref{TE}) can be solved numerically, where the ratio
$\Gamma(\theta)$ is independently measured in the DSMC method. All
power law exponents are found to be extremely accurate for all
driving mechanism, for all values of the restitution coefficient
$\alpha$, and for all dimensions (see Fig.2). So, a new class of
power law tails is obtained for WN and NF friction.

Let us now discuss the power law tail generated by an energy source
'at infinity', as recently proposed in Ref. \cite{EBN-Machta} with
suggestions for experimental tests. Here the source regularly
injects a macroscopic amount of energy in the system at ultra high
energies. This energy cascades down to lower energies, and builds
up a stationary 'low' energy tail $h(c)$. The mathematical
framework developed in the present paper is also well suited to
discuss this mechanism. By the arguments above (\ref{eq:Lambda})
the collision operator can be linearized, and the tail distribution
satisfies, $\Lambda h =0$. So we determine the right eigenfunction,
$h \sim c^{s^*-d-\nu}$ with eigenvalue $\lambda_{s^*} =0$. As
$\lambda_s( s
\to 0^+) = -1$, there does exist  a  non-vanishing positive solution
$s^*(d,\sigma,\alpha) >0$. As shown in Ref. \cite{appear}, the
equation for $s^*$ is equivalent to the one solved in
\cite{EBN-Machta} for the soft sphere model ($\sigma = \nu$). In
1-D it has the analytic solution $s^*(d=1)=1$ and $h(c) \sim
c^{-2-\nu}$. For $d>1$ the equation can been solved numerically.
Also the more traditional thermostats, as discussed in the context
of inelastic gases in Ref.\cite{Pias}, can be analyzed with the
help of the present method \cite{appear}.

In this Letter, we have studied a general class of inelastic soft
sphere models, on the basis of the nonlinear Boltzmann equation
coupled to stochastic or deterministic driving forces. We provide a
general framework which embeds in a versatile description  the
majority of driving devices and of particle interactions, discussed
in the literature: from hard scatterers like inelastic hard spheres
to soft scatterers like Maxwell molecules; from driving by white
noise, nonlinear friction, an energy source "at infinity", as well
as by traditional thermostats. It therefore represents a minimal
model for interpreting experiments on driven granular gases,
discussing trends, etc. When the non-equilibrium steady state is an
attractive fixed point of the dynamics, the velocity distribution
$f(c)$ has a stretched exponential tail $\propto\exp(-c^b)$;
sub-leading corrections appear in certain regions of model
parameters ($\nu,\alpha,\theta$), where $f(c)$ is found to be of
the form $c^\chi\exp(-\beta v^b+\beta' v^{b'})$. These corrections
turn out to be of particular relevance since they allow to obtain
an agreement with asymptotical theoretical predictions in a much
larger range of velocities, therefore more easily accessible to
experiments.  On the other hand, algebraic distributions emerge in
cases of marginal stability ($b=0$), where power law exponents have
been calculated. We have argued that such distributions should be
extremely difficult to observe experimentally.

Our theoretical starting point is backed up by recent experimental findings,
and all predictions --derived from a new analytical method-- have been tested
against very extensive Monte Carlo simulations \cite{appear}. Our work
therefore calls for further experimental investigations; in particular,
electrically forced gases with a charge distribution seem to provide
interesting test-beds.

\end{document}